\documentclass[global,twocolumn]{svjour}
\usepackage{latexsym}
\usepackage{graphicx,amssymb,subfigure}
\usepackage{amsmath,cuted}
\usepackage{hyperref}
\journalname{Applied Physics B}
\begin{document}
\title{All-Optical Single-Species Cesium Atomic Comagnetometer with Optical Free Induction Decay Detection}
\author{Yucheng Yang \and Teng Wu \and Jingbiao Chen \and Xiang Peng \and Hong Guo
}                     
\offprints{Hong Guo}          
\mail{hongguo@pku.edu.cn}
\institute{State Key Laboratory of Advanced Optical Communication Systems and Networks, Department of Electronics, and Center for Quantum Information Technology, Peking University, Beijing 100871, China}
\date{Received: date / Revised version: date}
%
\maketitle
\begin{abstract}
Atomic comagnetometers, which measure the spin precession frequencies of overlapped species simultaneously, are widely applied to search for exotic spin-dependent interactions.
Here we propose and implement an all-optical single-species Cs atomic comagnetometer based on the optical free induction decay (FID) signal of Cs atoms in hyperfine levels $F_g=3~\&~4$ within the same atomic ensemble.
We experimentally show that systematic errors induced by magnetic field gradients and laser fields are highly suppressed in the comagnetometer, but those induced by asynchronous optical pumping and drift of residual magnetic field in the shield dominate the uncertainty of the comagnetometer.
With this comagnetometer system, we set the constraint on the strength of spin-gravity coupling of the proton at a level of $10^{-18}$~eV, comparable to the most stringent one.
With further optimization in magnetic field stabilization and spin polarization, the systematic errors can be effectively suppressed, and signal-to-noise ratio (SNR) can be improved, promising to set more stringent constraints on spin-gravity interactions.
\end{abstract}
\section{Introduction}
\label{intro}
The spin-magnetic interaction is applied by atomic magnetometers to detect the magnetic field with a high sensitivity by measuring the atomic spin polarization precession frequency \cite{Budker2013}.
But to detect the non-magnetic spin-dependent interactions, the impact of magnetic field variation should be eliminated, and therefore the atomic comagnetometer scheme was presented \cite{Karwacki1980,Lamoreaux1986,Limes2018}.
By detecting the overlap of spin precession frequencies in the same magnetic environment, comagnetometers can suppress the impact of magnetic field variation in common mode, which may benefit the measurement of non-magnetic spin-dependent interactions in fundamental physics \cite{Safronova2018}, such as tests of CPT and Lorentz invariance \cite{Bear2000,Altarev2009,Allmendinger2014}, measurement of permanent electric dipole moments (EDMs) \cite{Griffith2009,Abel2017,Abel2020}, and searches for exotic spin-gravity interactions \cite{Kimball2013,Kimball2017,Wu2018,Mitchell2020,Wang2018}.

The magnetic field gradient is one of the main factors that lead to the systematic errors in comagnetometers.
Due to the difference in polarization \cite{Sheng2014}, gravity \cite{Baker2006} and/or thermal diffusion rate \cite{Ledbetter2012}, the species may have different average positions in the magnetic environment.
If the magnetic field gradients exist, different species may sense different magnetic fields, and the magnetic field variations in common mode can not be suppressed.
Therefore, comagnetometers with various species will suffer from the accuracy reduction from the magnetic field gradients.

Measures are taken to fix the errors induced by magnetic field gradients, such as monitoring the Larmor frequency shift as a function of the applied magnetic field gradients \cite{Afach2015}, or correcting for the magnetic field gradients via theoretical calculations \cite{Sheng2014}.
But these numerical methods are dependent on other parameters, such as the size of the atomic vapor cells, and may bring calibration errors.
The scheme adopting alkali metal atoms ($\rm ^{85}Rb~\&~^{87}Rb$ \cite{Kimball2013,Kimball2017}) can seemingly almost root out the influence from the magnetic field gradients, because the fast diffusion rates of gas atoms will lead to almost same average positions.
But the frequency shift induced by the magnetic field gradients \cite{Cates1988}, which is related to the gyromagnetic ratio \cite{Pendlebury2004}, can still degrade the performance of the $\rm ^{85}Rb~\&~^{87}Rb$ comagnetometer.
To eliminate the systematic errors from magnetic field gradients, the single-species scheme was proposed in a nuclear-spin comagnetometer based on a liquid of identical molecules \cite{Wu2018}, and the comagnetometer was experimentally shown to have suppressed systematic errors from magnetic field gradients.
The performance of the nuclear-spin comagnetometer is limited by the small polarization ratio of the nuclear spins.

Laser fields also produce systematic errors in comagnetometers.
The spin-precession frequencies of atoms will be shifted by the light (the light shift effect \cite{Kastler1963}), which accounts for the shifted result of the measured magnetic field.
Furthermore, atoms illuminated by the laser light will have larger relaxation rate because of power broadening, which will degrade the systematic sensitivity.
Consequently, laser light may deteriorate the accuracy and sensitivity of each magnetometer, and the errors caused by pump light and probe light should be calibrated carefully \cite{Kimball2013,Kimball2017}.

In this paper, we put forward and carry out the single-species comagnetometer scheme making use of free induction decay (FID) signal of atomic spin polarization in an all-optical \cite{Bell1961} nonlinear magneto-optical rotation (NMOR) \cite{Budker2002} Cs magnetometer.
Compared with comagnetometers with different overlapped species \cite{Sheng2014,Ledbetter2012,Afach2015,Kimball2013,Kimball2017}, our single-species atomic comagnetometer can suppress the systematic errors induced by magnetic field gradients, considering the fast atomic diffusion rate and the almost identical gyromagnetic ratios of the two hyperfine levels $F_g=3~\&~4$.
Compared with the nuclear-spin comagnetometer \cite{Wu2018,Wu2019}, our atomic comagnetometer has higher spin polarization ratio, and is promising to have a better signal-to-noise ratio (SNR).

A similar $\rm ^{87}Rb$ atomic comagnetometer has been implemented by our group \cite{Wang2018}. As is shown in Table\ref{tab:alkaliatoms}, Cs atoms are preferred for 3 reasons:\\
(1)~The difference in gyromagnetic ratio of hyperfine levels in the ground state $\Delta\gamma$ is large, which means that, the required magnetic field to make the MR signals resolvable is small.\\
(2)~The splitting of hyperfine levels in the ground state $\Delta\nu_{\rm HFS}$ is the largest in alkali metal atoms commonly used for magnetometers, which corresponds to the least nonlinear Zeeman shifts of magnetic sublevels in the same applied magnetic field, according to the Breit-Rabi formula \cite{Breit1931}.\\
(3)~The vapor pressure $P_V$ at room temperature (298 K) is the largest, which will result in the largest MR signal amplitudes and best systematic SNR in the same temperature.
\begin{table}
\caption{Some Data of Alkali Metal Atoms Commonly Used for Magnetometers}
\label{tab:alkaliatoms}      
\begin{tabular}{ccccc}
\hline\noalign{\smallskip}
~ & K~\cite{K} & $\rm ^{85}Rb$~\cite{Rb85}& $\rm ^{87}Rb$~\cite{Rb87}& Cs~\cite{Cs}\\
\noalign{\smallskip}\hline\noalign{\smallskip}
$\Delta\gamma \rm{(mHz/nT)}$ & 3.973 & 8.220 & 27.857 & 11.165\\
$\Delta \nu_{\rm HFS}\rm{(GHz)}$ & 0.461 & 3.035 & 6.835 & 9.192\\
$P_V @298~\rm{K(Torr)}$ & $10^{-7.6}$ & $10^{-6.4}$ & $10^{-6.4}$ & $10^{-5.8}$\\
\noalign{\smallskip}\hline
\end{tabular}
\end{table}

Recently, we found that a dual frequency Cs spin maser of a similar scheme was presented \cite{Bevington2019}, in which the authors experimentally show that the systematic errors from light field still exist.
In our system, the influence of the pump light, such as light shift and cross talk between hyperfine Zeeman sublevels, is almost eradicated because the pump light is blocked when detecting the optical free induction decay (FID) signal.
\section{Experimental scheme}
\begin{figure*}
\resizebox{\textwidth}{!}{
\includegraphics{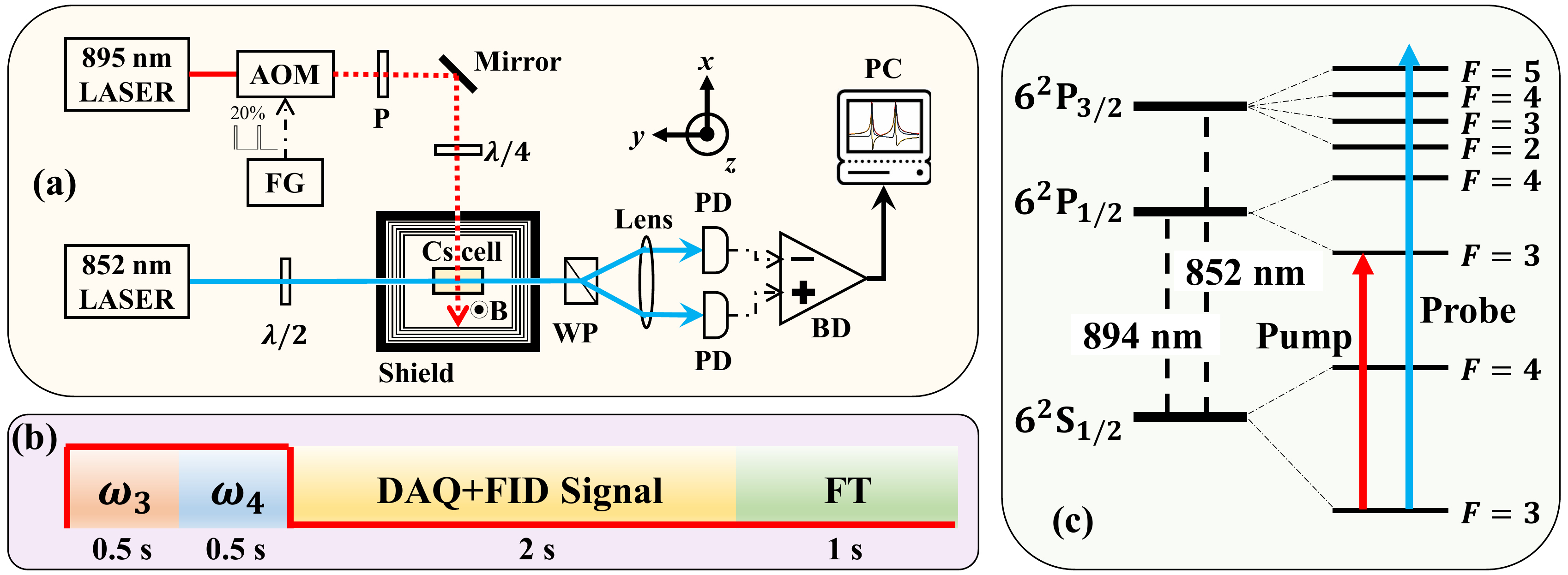}}
\caption{Experimental scheme of the single-species comagnetometer system. 
\textbf{a} The apparatus applied in experiments. The pump light (red solid line) is emitted from an 895 nm (Cs D1 line) Laser. Its amplitude is modulated by a square wave (20$\%$ duty cycle). The modulated pump light (red dotted arrow) later passes through a quarter wave plate and becomes circularly polarized to illuminate the Cs vapor cell.
The linearly polarized probe light (blue solid arrow) is generated from an 852 nm (Cs D2 line) Laser. Its polarization is rotated in the Cs vapor cell, and the rotation signal will be recorded by a balanced detector and later analyzed in a processor computer.
The Cs vapor cell with anti-relaxation coatings is placed in a 7-layer magnetic shield, immersed in a static magnetic field generated by a constant current source.
AOM: acousto-optic modulator, $\lambda/2$: half wave plate, $\lambda/4$: quarter wave plate, BD: balanced detector, P: polarizer, PC: processor computer, PD: photodiode, FG: function generator, WP: Wollaston prism, FID Signal: free induction decay signal, MR Signal: magnetic resonance signal.
\textbf{b} Steps of one run in the system. Step 1: the pump light is turned on and has its amplitude modified at Larmor frequency of $F_g=4$ for 0.5 s; Step 2: the amplitude of the pump light is modulated at Larmor frequency of $F_g=3$ for 0.5 s; Step 3: the pump light is turned off and the data acquisition (DAQ) starts to record the FID signal (2 s); Step 4: Recorded FID signal is transformed into MR signals in spectrum via Fourier transformation and later fitted in the processor computer (1 s, this step can be done offline). 
\textbf{c} Energy levels of Cs D1 and D2 transition (not to scale). The pump light (red arrow) is tuned to the center of the Doppler-broadened Cs D1 $F_g=3\rightarrow F_e=3$ resonance, and the probe light (blue arrow) is 5 GHz blue-detuned from the center frequency of the Doppler-broadened Cs D2 $F_g=3\rightarrow F_e=4$ resonance.
}
\label{fig:setup}
\end{figure*}
\label{sec2}
The experimental setup is shown in Fig. \ref{fig:setup}a.
At the center of our system is a self-made paraffin coated cylinder Cs vapor cell (diameter = 2.5 cm, length = 2.5 cm), which was manufactured by ourselves. The longitudinal spin relaxation time of the Cs vapor cell was measured to be $T_1\approx 3.3~\rm{s}$, which is limited by ``uniform relaxation'' \cite{Li2017}, i.e., the exchange of alkali atoms between the volume and the stem. The anti-relaxation coatings in the inner surface can effectively reduce the relaxation due to wall collisions. The Cs vapor cell is located in a seven-layer magnetic shield (manufactured by Beijing Zero-Magnet Technology Co., Ltd) made of a 1-mm thick high-permeability alloy. The temperature of the shield is stabilized at $\rm 22~^{\circ}C$ to provide the Cs vapor cell a stable temperature environment, yielding vapor density of $n \approx 3.5\times 10^{10}~\rm{atoms/cm^{3}}$. Within the shield is a set of three-dimensional Helmholtz coils, driven with a current source (Krohn-hite Model 523 calibrator, stability $\pm 1~$ppm within 24 hours) to generate a bias DC magnetic field.

In order to measure the spin precession frequency of Cs atoms in $F_g=3~\&~4$, there are two main processes in this comagnetometer -- preparation and measurement of atomic spin polarization (see Fig. \ref{fig:setup}b).

During the preparation of atomic spin polarization (step $1~\&~2$ in Fig. \ref{fig:setup}b, duration = 1 s), Cs atoms are illuminated with a left-circularly polarized pump light propagating along $-\hat{x}$ (orthogonal to $\bf B$, which is along $\hat{z}$) tuned to the center of the Doppler-broadened Cs D1 $F_g=3\rightarrow F_e=3$ resonance, where $F_g=3$ is the ground-state hyperfine level and $F_e=3$ is the excited-state hyperfine level (see the red arrow in Fig. \ref{fig:setup}c).
The 895~nm D1 pump beam is generated with a distributed Bragg reflector laser diode (Photodigm PH-895-DBR-080-T8). The peak value of the pump light power is $\approx 3.75$ mW, and the beam size is $\rm \approx 4~mm^{2}$.
In the pumping process, most atoms in $F_g=3$ are depopulated by the pump light, except for atoms in Zeeman sublevel $m=+3$; while because the excited atoms will repopulate to all Zeeman sublevels except for $m=-4$ in $F_g=4$, the 1-order polarizations (orientations) in both hyperfine levels $F_g=3~\&~4$ are generated.
To polarize the spin in both hyperfine levels, the pump light has its amplitude modulated at Larmor frequency of $F_g = 4~\&~3$ successively with an acousto-optic modulator (AOM, ISOMET M1250-T150L-0.5).
The duty cycle of the modulation is chosen at 20\% to maximize the spin polarization with a relatively long transverse spin relaxation time $T_2$ \cite{Gerginov2017}.
Note that, to make the MR signal amplitudes of $F_g=3~\&~4$ comparable, the spin in $F_g=4$ must be polarized in step 1, and later to polarize the spin in $F_g=3$ in step 2; otherwise, because of the faster relaxation rate and lower polarization ratio for atoms in $F_g=3$, the MR signal amplitude of $F_g=3$ is about one order of magnitude smaller than that in $F_g=4$.

During the measurement of atomic spin polarization (step 3 in Fig. \ref{fig:setup}b, duration = 2 s, longer than the spin polarization relaxation time, which is about 106~ms for $F_g=4$ and 53~ms for $F_g=3$ in our experiments), the pump beam is blocked with the AOM, and a linearly polarized probe beam propagating along $-\hat{y}$ travels through the Cs vapor cell and into a polarimeter. The 852~nm D2 probe beam is produced with a tunable external-cavity diode laser (New Focus TLB-6817) with an isolator (Thorlabs IO-3D-850-VLP, not shown in Fig. \ref{fig:setup}a). The spin precession frequencies of atoms in $F_g=3~\&~4$ are measured by observing optical rotation of the probe light. The probe light is 5 GHz blue-detuned from the center frequency of the Doppler-broadened Cs D2 $F_g=3\rightarrow F_e=4$ resonance (see the blue arrow in Fig. \ref{fig:setup}c), the power is $\approx 1.6$ mW, and the beam size is $\rm \approx 4~mm^{2}$. The probe beam is split with a Wollaston prism and detected with a balanced detector (Thorlabs PDB210A). The optical FID signal from the balanced detector is sampled with a multifunction I/O device (National Instruments USB-6363) using a routine written in LabVIEW.

After the data acquisition, the MR signals are acquired through the Fourier transformation of the FID signal. And the data analysis (step 4 in Fig. \ref{fig:setup}b, duration = 1 s) is conducted by fitting the MR signals to an overall Lorentzian profile:
\begin{equation}
L(f)=\sqrt{\left[\sum\frac{A_F}{1+\left(\frac{f-f_F}{\nu_F}\right)^2}\right]^2+\left[\sum\frac{B_F(f-f_F)/\nu_F}{1+\left(\frac{f-f_F}{\nu_F}\right)^2}\right]^2},
\label{eq:Rsignal}
\end{equation}
where $L(f)$ is the theoretical form of the overlapped MR signals in spectrum, $A_F$ and $B_F$ represent the amplitudes of the imaginary and real part of the signals, respectively, $f_F$ is the corresponding Larmor frequency of corresponding hyperfine level $F$, and $\nu_F$ is the resonance width (HWHM) of the MR signal.
All steps take 4 seconds in total.

To measure the non-magnetic spin-dependent interactions, comagnetometers should operate under proper magnetic field $B_0$, which makes the two MR signals resolvable in the spectrum and meanwhile the broadening in the widths of MR signals due to nonlinear Zeeman effect is not obvious. 
Improper magnetic field may degrade the accuracy and/or sensitivity of each magnetometer in the comagnetometer system, thus deteriorate the potential to set more stringent constraints on exotic spin-dependent couplings.

\begin{figure}
\resizebox{\columnwidth}{!}{
\includegraphics{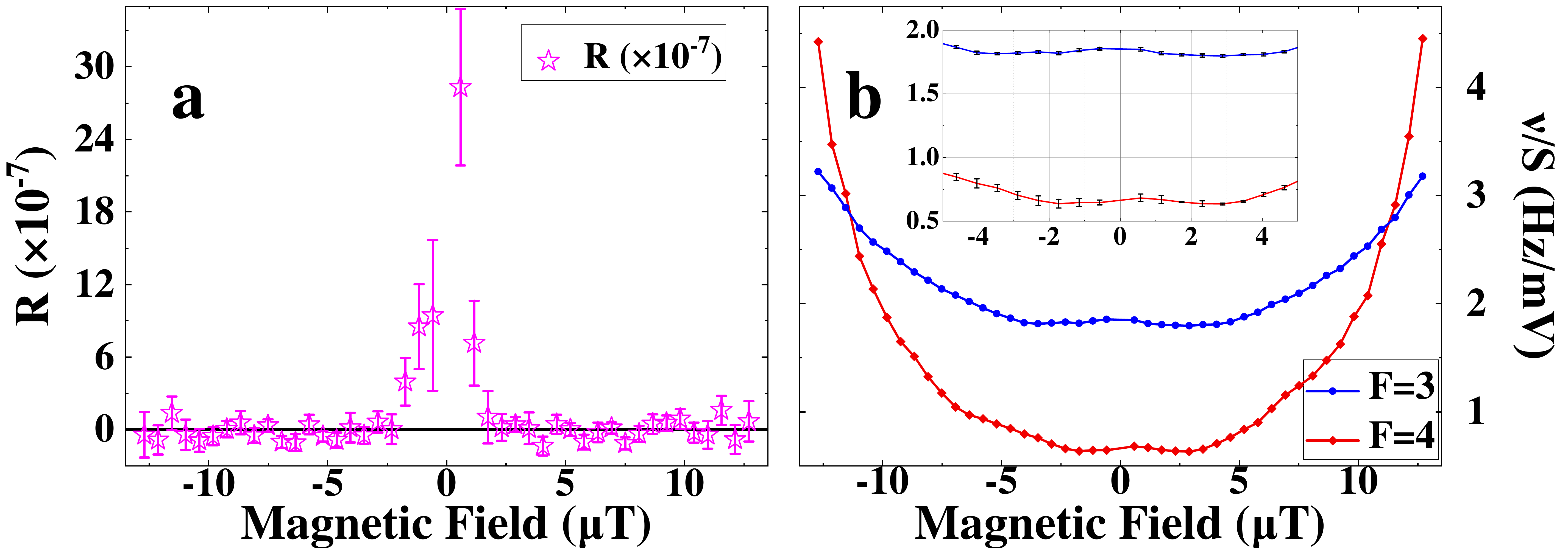}}
\caption{\textbf{a} The dependence of the index $R$ (deviation of measured frequency ratio from theoretical value) on the applied magnetic field $B_0$. When $B_0 < 2~{\rm \mu T}$, the two MR signals are so close to each other in the spectrum that the fitting error will limit the accuracy on frequency measurement, leading to the large $R$. For $B_0\ge 2~{\rm \mu T}$, $R$ fluctuates within $\pm 4\times 10^{-7}$. \textbf{b} The dependence of the width-to-amplitude ratio $\nu/{\rm S}$ on the applied magnetic field $B_0$. The parameter $\nu/{\rm S}$ is proportional to the sensitivity of frequency measurement. The inset shows the rescaled results for $B_0 < 5~{\rm \mu T}$, indicating that within this range the measurements for Larmor frequencies of both hyperfine levels have good sensitivity. In our system, the optimal range of the applied magnetic field $B_0$ should be from $2~{\rm \mu T}$ to $5~{\rm \mu T}$.}
\label{fig:index}
\end{figure}

When the $B_0$ is small, the two MR signals of hyperfine levels in the ground state are not resolvable in the frequency spectrum, increasing the fitting error.
Bad fitting may account for the error of the measured magnetic field and therefore, the failure in suppressing the magnetic field variations.
To judge the accuracy of the two magnetometers, we construct the index $R$, the deviation between the measured frequency ratio and the theretical value, as
\begin{equation}
R=\frac{f_{3}}{f_{4}}/\frac{\gamma_{3}}{\gamma_{4}}-1,~ \frac{\gamma_{3}}{\gamma_{4}}=\frac{g_j-9g_I}{g_j+7g_I}\approx 1.003191233,
\label{eq:definer}
\end{equation}
where $f_{3,4}$ are the measured center frequencies of MR signals for corresponding hyperfine levels, $\gamma_{3,4}$ are the gyromagnetic ratios, $g_{j,I}$ is the Land$\rm \acute{e}$ factor of electron and nuclei, and $|\gamma_{3}/\gamma_{4}|$ can be calculated from Ref. \cite{Bergmann1997}.
As is shown in Fig. \ref{fig:index}a, when the applied magnetic field $B_0 < 2~{\rm \mu T}$, $R$ is too large, the accuracy of each magnetometer is limited by the fitting error. And when $B_0>2~{\rm \mu T}$, the comagnetometer performance is immune to the magnetic field variations.
When the $B_0$ is large, the widths of the two MR signals will be broadened due to nonlinear Zeeman effect, and the sensitivity of the comagnetometer to detect non-magnetic spin-dependent interactions will be worsened.
The sensitivity of the comagnetometer for measuring non-magnetic spin-dependent interaction is determined by the sensitivity of the two magnetometers that constitute the comagnetometer, and the sensitivity of a magnetometer is
\begin{equation}
\delta B\propto \frac{\nu}{\gamma\cdot\rm{S/N}},
\label{eq:sens}
\end{equation}
where $\gamma$ is the gyromagnetic ratio of working atoms, $\nu$ is the MR width (HWHM), and S/N is the signal-to-noise ratio of the system. However, due to nonlinear Zeeman effect, the amplitude becomes reduced, and the width becomes broadened, when the magnetic field $B_0$ grows.
In the system, the noise is relatively constant, and the sensitivity of the magnetometer, which is depicted with Eq. (\ref{eq:sens}), is mainly determined by the ratio of MR signal width and signal amplitude $\nu/{\rm S}$.
The dependence of this ratio on the magnetic field strength is shown in Fig. \ref{fig:index}b, and the inset figure indicates that when the applied magnetic field $B_0 < 5~{\rm \mu T}$, both magnetometers of $F_g=3~\&~4$  have good sensitivity.

Overall, to achieve the optimal accuracy and sensitivity, the Cs comagnetometer should work at a magnetic field $B_0$ ranging from 2~$\mu$T to 5~$\mu$T.
\section{Measurement and data analysis}
\label{sec:3}
\subsection{Estimation for systematic errors}
\label{subsec:1}
For the single-species comagnetometer system, the index $R$ is related to the ability to search for non-magnetic spin-dependent interactions. According to Eq. (\ref{eq:definer}), the systematic errors in measuring frequencies will finaly effect on the index $R$ as
\begin{equation}
R_{\rm err}=\frac{1}{f_4}\sqrt{\left(\frac{\gamma_4}{\gamma_3}\right)^2 f_{\rm 3,err}^2+f_{\rm 4,err}^2},
\label{eq:Rerror}
\end{equation}
where $f_{\rm 3/4,err}$ are the corresponding systematic errors in measuring the Larmor frequencies of hyperfine levels.

The sources of systematic errors in the system can be divided into four parts -- the magnetic field, the (pump and probe) light fields, the atomic collisions, and the Earth rotation.
In this section, the estimations for systematic errors from each part are presented.
\label{analysis}
\subsubsection{Magnetic field}
\label{subsubsec1}
We verified that our atomic comagnetometer system is immune to the magnetic field gradients, by measuring the index $R$, which is depicted with Eq. (\ref{eq:definer}).
Fig. \ref{fig:index}a also shows that, the systematic errors from the magnetic field gradients, which are common in traditional comagnetometers, are suppressed in our Cs comagnetometer.
Assuming the difference in magnetic field experienced by atoms in $F_g=3~\&~4$ is $\Delta B$, the index $R$ is given by
\begin{equation}
R\approx\frac{\gamma_{3}(B+\Delta B)}{\gamma_{4}B}/\frac{\gamma_{3}}{\gamma_{4}}-1=\frac{\Delta B}{B}.
\label{eq:gradient}
\end{equation}
If the Cs comagnetometer is sensitive to the magnetic field gradients, $R$ should be proportional to $B^{-1}$. However, Fig. \ref{fig:index}a shows that, at the level of the measurement uncertainty, the measured $R$ has almost no dependence on the strength of the bias field $B$.

The geometric phase effect \cite{Pendlebury2004}, which is related to the magnetic field gradients, is highly suppressed in our system, too. The frequency shift due to the geometric phase effect is given by \cite{Pendlebury2004}
\begin{equation}
\Delta f_{F,\rm geo}=\frac{(\gamma_F B_{r})^2}{2}\frac{\gamma_F B}{(\gamma_F B)^2-f_r^2},
\label{eq:geo}
\end{equation}
where $B_{r}=\nabla B\cdot r/2$, $f_r=\bar{v}/2\pi r$, $\nabla B$ is the magnetic field gradients, $\bar{v}$ is the mean thermal velocity of atoms, and $r$ is the radius of the cell.
In our system, the applied magnetic field is $B=3463.8~\rm nT$, the cell radius is $r=1.25~\rm cm$, the mean thermal velocity of Cs atoms at $22^\circ\rm~C$ is $\bar{v}=216.73~\rm m/s$, and the magnetic field gradient along the direction of $B$ is $\nabla B_z=1.5~\rm nT/cm$.
Substitute these parameters into Eq. (\ref{eq:geo}), and the frequency shifts due to the geometric phase effect are
\begin{equation}
\begin{aligned}
\Delta f_{3,\rm geo}&=-4.59\times 10^{-4}~\rm Hz,\\
\Delta f_{4,\rm geo}&=-4.52\times 10^{-4}~\rm Hz.
\label{geoshift}
\end{aligned}
\end{equation}
Combining Eq. (\ref{eq:Rerror}) and Eq. (\ref{geoshift}), the upper limit of the systematic errors induced by the geometric phase effect in the index $R$ is calculated to be
\begin{equation}
R_{\rm err,geo}=5.325\times 10^{-8}.
\end{equation}

The nonlinear Zeeman effect is another way for the magnetic field to cause systematic errors in the comagnetometer system. The energy shift of each Zeeman sublevels $|F,m_F\rangle$ for Cs atoms at the magnetic field $B$ is given by the Breit-Rabi formula \cite{Breit1931} as
\begin{equation}
\begin{aligned}
\Delta E(F,m_F)=g_I\mu_B m_F B&\pm\frac{\Delta E_{\rm HFS}}{2}\left(1+\frac{m_F x}{2}+x^2\right)^{1/2},\\
x=&\frac{(g_j-g_I)\mu_B B}{\Delta E_{\rm HFS}},
\label{eq:Zeemanlevels}
\end{aligned}
\end{equation}
where $\mu_B$ is the Bohr magneton, $\Delta E_{\rm HFS}=h\cdot\Delta\nu_{\rm HFS}$ is the energy splitting between ground-state hyperfine levels, $\pm$ corresponds to the hyperfine levels $F=I\pm j$. The spin precession frequency of atoms in Zeeman sublevel $|F,m_F\rangle$ is given by
\begin{equation}
f(F,m_F)=\frac{\Delta E(F,m_F)-\Delta E(F,m_{F}-1)}{h},
\end{equation}
and the MR signal of hyperfine level $F$ can be treated as the overlap of $2F$ MR signals with different frequency.
In our system, because the pump light is left-circularly polarized and tuned to the center of $F_g=3\rightarrow F_e=3$ transition, atoms in $|F_g=3,m_F=-3\sim 2\rangle$ will be depopulated to $|F_e=3, m_F=-2\sim 3\rangle$. The excited atoms in $|F_e=3, m_F=-2\sim 3\rangle$ will spontaneously transit back to the Zeeman sublevels $|F_g=3, m_F\rangle$ and $|F_g=4, m_F=-3\sim4\rangle$in the ground state. Atoms repopulating in $F_g=3$ will be depopulated again. In the steady state, atoms in $F_g=3$ mostly populate in $m_F= 3$, and the MR frequency is given by
\begin{equation}
f_{\rm 3,NLZ}=\frac{\Delta E(3,3)-\Delta E(3,2)}{h}.
\label{eq:nlz3}
\end{equation} 
The situation in $F_g=4$ is more complicated. Due to the same transition rate for repopulation $|F_e=3, \pm m_F\rangle\rightarrow|F_g=4\rangle$, the contribution of atoms repopulating in $F_g=4$ from $|F_e=3,m_F=-2\sim2\rangle$ to overall MR signal is ``neutralized''. The max frequency shift for $F_g =4$ due to nonlinear Zeeman effect is given in the approximation that, only atoms repopulating in $F_g=4$ via the transition $|F_e=3, m_F=3\rangle\rightarrow|F_g=4,m_F=2,3,4\rangle$ contribute to the comprehensive MR signal of $F_g=4$, as
\begin{equation}
f_{\rm 4,NLZ}=\frac{\Delta E(4,4)-\Delta E(4,1)}{3h}.
\label{eq:nlz4}
\end{equation} 
Substituting the applied magnetic field is $B_0=3463.8~\rm nT$ in our system into Eq. {\ref{eq:definer}}, Eq. (\ref{eq:nlz3}) and Eq. (\ref{eq:nlz4}), the upper limit of the systematic errors induced by nonlinear Zeeman effect is calculated to be
\begin{equation}
R_{\rm err,NLZ}=2.13\times 10^{-8}.
\end{equation}
The result is one order in magnitude larger than that in Ref. \cite{Wang2018}. This may be because that the alignment is measured in Ref. \cite{Wang2018}, and the nonlinear Zeeman effect is better suppressed in the comprehensive MR signal by averaging the MR signals of atoms in $|F_g,\pm m_F\rangle$.

\subsubsection{Pump light}
\label{sectionpumplight}
Our FID comagnetometer is proved capable of highly suppressing power broadening and light shift effect from the pump light, compared with single-species comagnetometers which obtain the MR signals by scanning the modulation frequency or magnetic field \cite{Wang2018,Bevington2019}.
The comparison is conducted by scanning the pump light power in FID mode and modulation-frequency-scan (MFS) mode at the same magnetic field $B_0=$3463.8 nT, using the same equipment. The information of the signal from polarimeter in MFS mode is acquired with a lock-in-amplifier (Stanford Research SR850) and later collected by a LabVIEW routine.
\begin{figure}
\centering
\resizebox{\columnwidth}{!}{%
\includegraphics{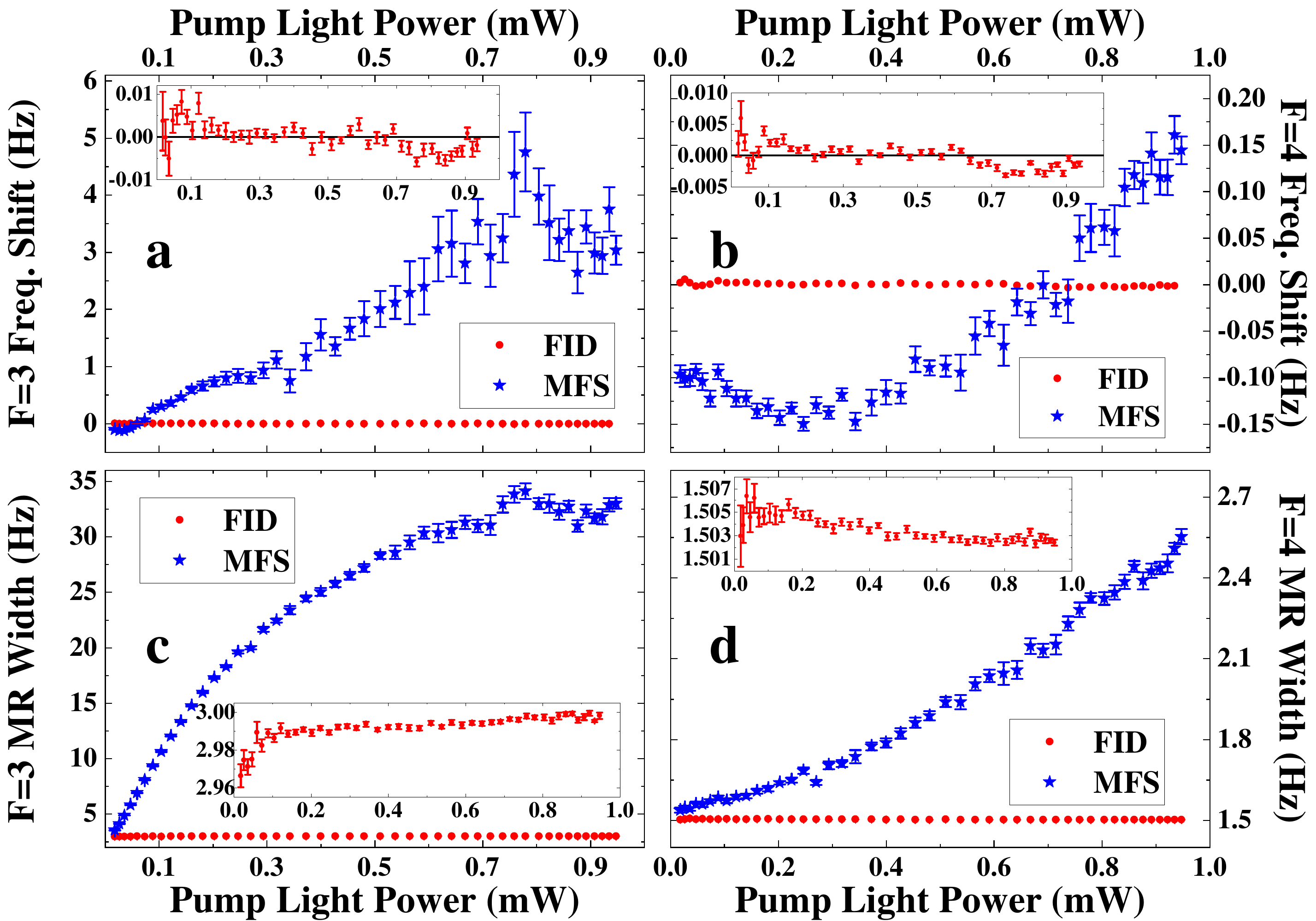}}
\caption{\textbf{a} The dependence of the measured shift in Larmor frequency of $F_g=3$ ($f_3=12155.838~{\rm Hz}$ in theory when $B_0=3463.8~{\rm nT}$) on the pump light power for FID mode (red dot) and MFS mode (blue star).
\textbf{b} The dependence of the measured shift in Larmor frequency of $F_g=4$ ($f_4=12117.169~{\rm Hz}$ in theory when $B_0=3463.8~{\rm nT}$) on the pump light power for FID mode (red dot) and MFS mode (blue star). 
\textbf{c} The dependence of the measured MR width (HWHM) of $F_g=3$ on the pump light power for FID mode (red dot) and MFS mode (blue star). 
\textbf{d} The dependence of the measured MR width (HWHM) of $F_g=4$ on the pump light power for FID mode (red dot) and MFS mode (blue star). 
The insets in each figure are the rescaled corresponding results in FID mode. Due to inadequate pumping rate when the pump light power is small, the errors of the four variables are large.
The light shift effect and power broadening are highly suppressed in FID mode.}
\label{fig:power}
\end{figure}

The atomic spin precession frequency, in the MFS mode, varies with the pump light power, because of the light shift effect in both $F_g=3$ (shown in Fig. \ref{fig:power}a) and $F_g=4$ (shown in Fig. \ref{fig:power}b).
Within the range of the pump light power, the atomic spin precession frequency varies within a range of 6 Hz in $F_g=3$, and 0.4 Hz in $F_g=4$.
The difference may come from the light configuration shown in Fig. \ref{fig:setup}c, which implies that the vector light shift is more obvious in $F_g=3$ due to the pump light.
While in the FID mode, the atomic precession frequencies of both hyperfine levels fluctuate within $\rm0.01~Hz$ when the pump light power is more than 0.2 mW.

In regard to the MR width (HWHM), in the MFS mode, it grows with the pump light power due to the power broadening effect in $F_g=3$ (shown in Fig. \ref{fig:power}c) and $F_g=4$ (shown in Fig. \ref{fig:power}d).
With the increase of the pump light power, the MR width of $F_g=3$ grow steeply (31~Hz broadened at max pump light power), because the pump light interacts with atoms in $F_g=3$ directly;
and the MR width of $F_g=4$ also increases slightly (1~Hz broadened at max pump light power), due to the Doppler broadening effect and the natural width of atoms.
While in FID mode, the MR widths of both hyperfine levels are almost uniform (2.993 Hz for $F_g=3$, pink line; and 1.503 Hz for $F_g=4$, green line).
The width difference mainly comes from the inconsistent relaxation rate induced by spin-exchange effect \cite{Happer1977}.
The broadening in MR signal width $\Delta\nu_{\pm}$ due to spin-exchange collisions can be calculated as:
\begin{equation}
\Delta\nu_{\pm}=\frac{\Gamma_{\pm}}{2\pi}\approx\frac{1}{2\pi}\left(\frac{K\mp 1}{T_{SE}\cdot I}\pm\frac{K^2-1}{2\omega_F^{2}T^{3}I^{3}}\right),~K=\frac{I^2+2}{3I},
\label{eq:spinexch}
\end{equation}
where $\Gamma_{\pm}$ is the relaxation rate induced by spin exchange collisions, corresponds to the broadening in MR signals of hyperfine levels $F_g=I\pm1/2$, $I$ is the nuclear spin (for Cs, $I=7/2$), $T_{SE}$ is the spin-exchange time calculated by $T_{SE}=(n\sigma_{\rm SE}\bar{v})^{-1}$, $\sigma_{\rm SE}$ is the cross section of spin-exchange collisions between Cs atoms, and $\bar{v}$ is the average thermal velocity of Cs atoms.
In our experiments, the cell temperature is stabilized at $T=22~^{\circ}\rm{C}$, and the broadening due to spin-exchange collisions are:
\begin{equation}
\Delta\nu_{3}\approx 1.73~\rm{Hz},~~~\Delta\nu_{4}\approx 0.26~\rm{Hz},
\label{eq:width}
\end{equation}
the difference is 1.47 Hz in theory, which coincides well with the measured $\sim 1.33$~Hz difference in the width of the two MR signals in FID mode.

\begin{figure}
\centering
\subfigure{
\begin{minipage}[t]{\columnwidth}
\centering
\includegraphics[width=0.6\columnwidth]{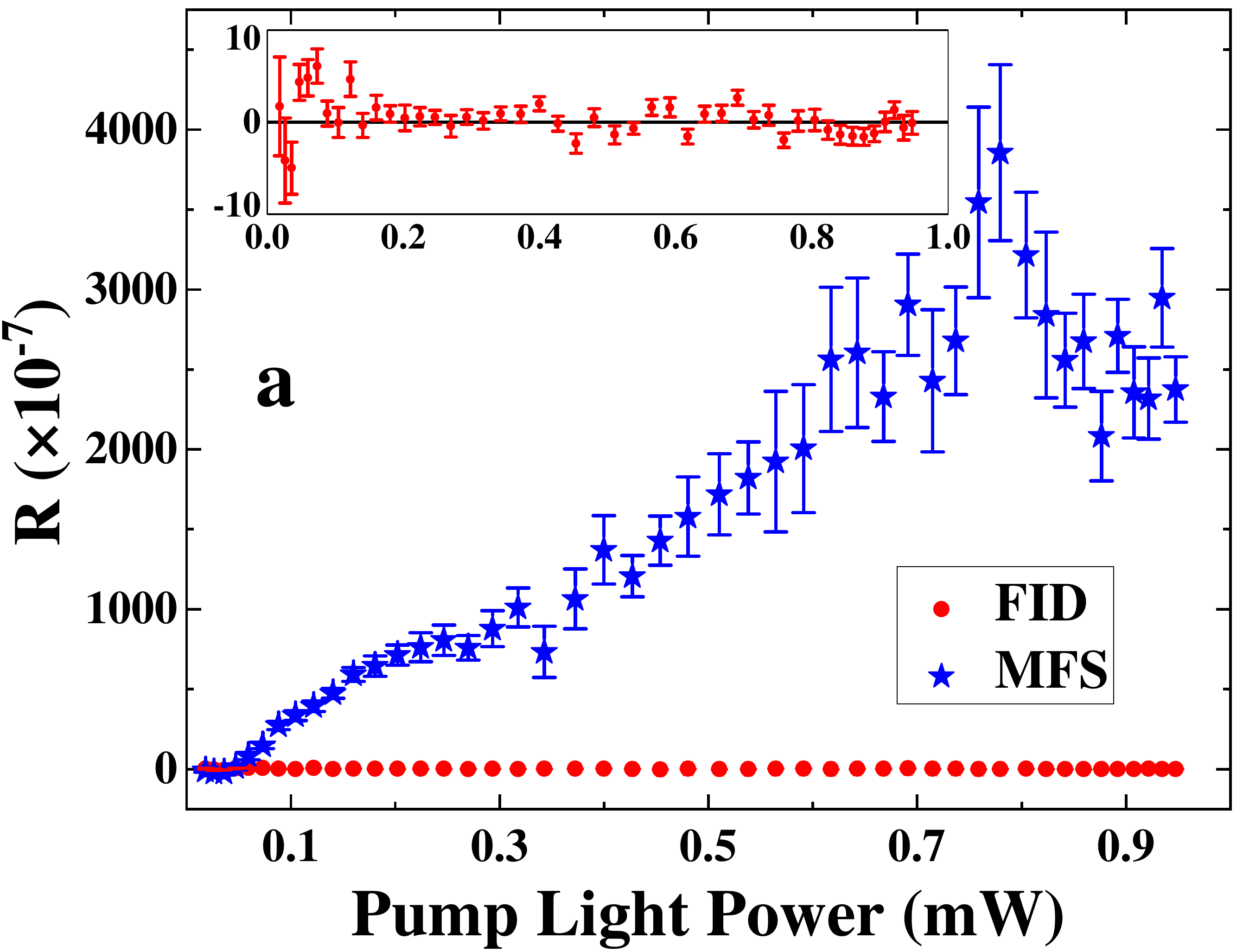}
\end{minipage}%
}%
\\
\subfigure{
\begin{minipage}[t]{\columnwidth}
\centering
\includegraphics[width=0.6\columnwidth]{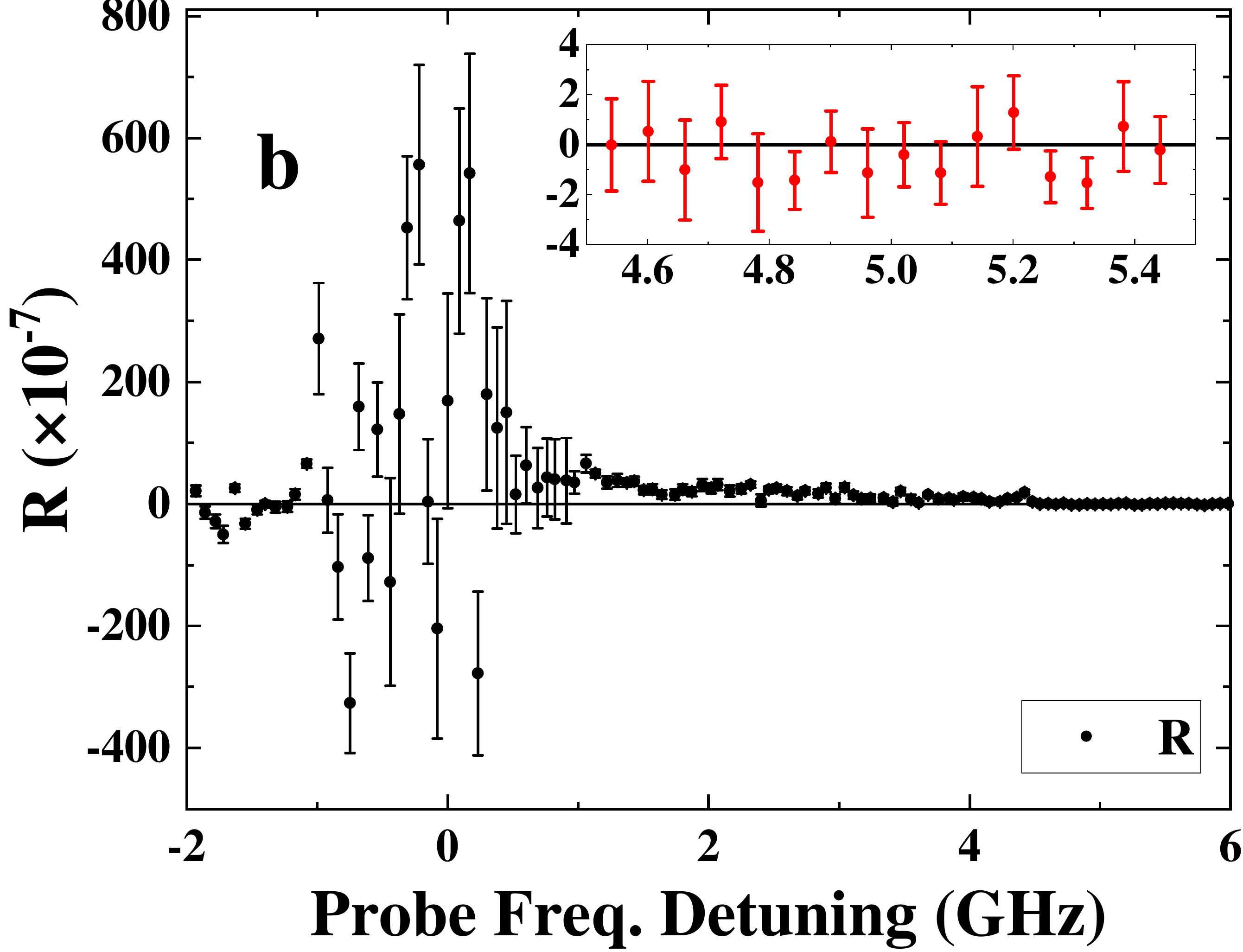}
\end{minipage}%
}%
\\
\subfigure{
\begin{minipage}[t]{\columnwidth}
\centering
\includegraphics[width=0.6\columnwidth]{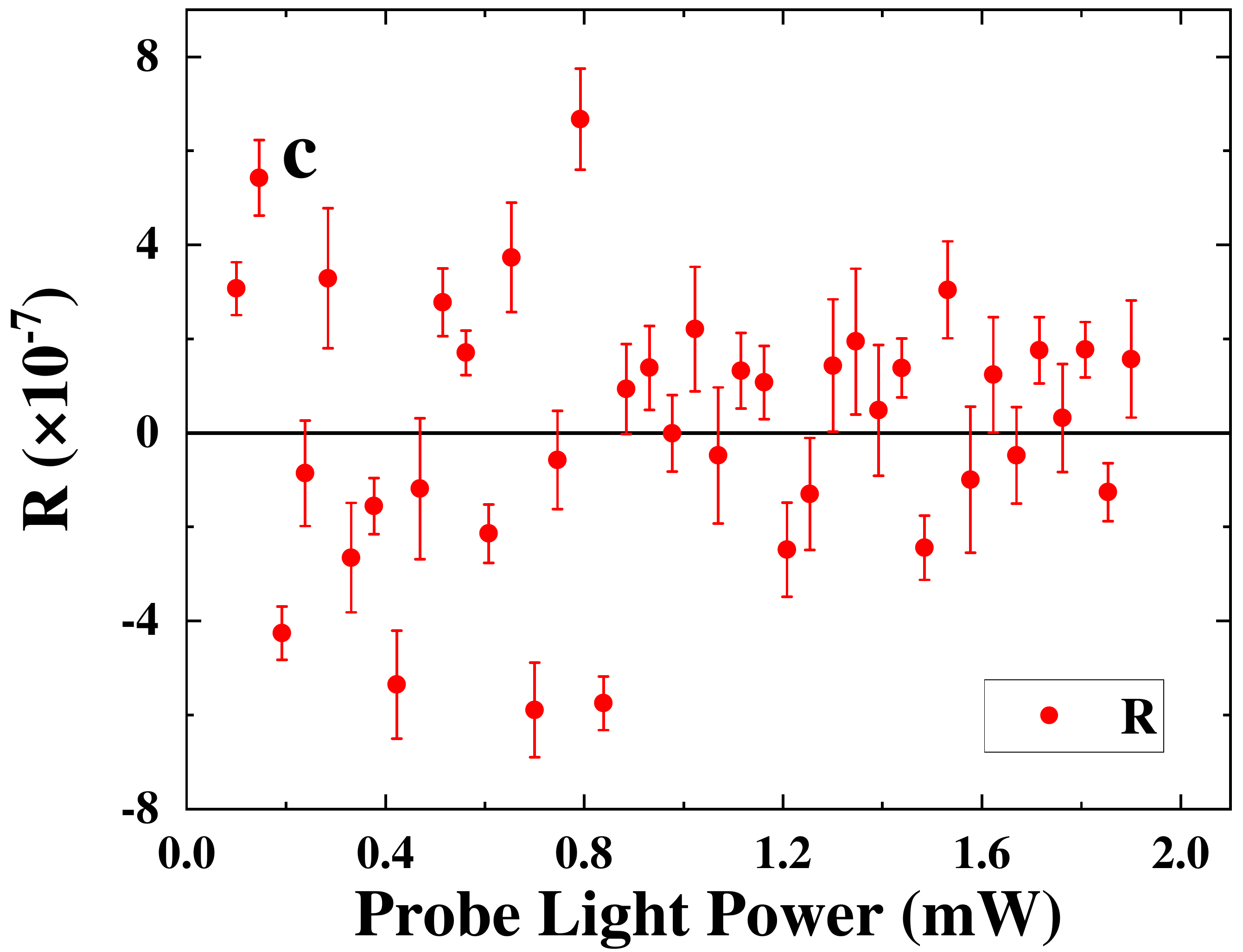}
\end{minipage}
}%
\centering
\caption{\textbf{a} The dependence of $R$ on the pump light power in FID mode (red dot) and MFS mode (blue star). The inset shows the rescaled results in FID mode. $R$ in MFS mode can be 200 times (4000:20) larger than that in FID mode.
\textbf{b} The dependence of $R$ on the frequency detuning the probe light from Cs D2 $F_g=3\rightarrow F_e=4$ transition (set as 0). When the probe light is near-resonance from Cs D2 $F_g=3\rightarrow F_e$ transition ($-1\sim 1$ GHz), the deviation $R$ is very large due to the pumping effect of the probe light. The probe light power is fixed at 1.6 mW. The inset shows the rescaled results (red dot) when the detuning is from 4.5 GHz to 5.5 GHz.
\textbf{c} The dependence of $R$ on the probe light power. The probe light is 5 GHz blue-detuned from the center frequency of the Doppler-broadened Cs D2 $F_g = 3 \rightarrow F_e = 4$ resonance.
The index $R$ will fluctuate within $\pm 4\times 10^{-7}$ under typical operation conditions in our experiments.
}
\label{fig:Rdependence}
\end{figure}

Due to the power broadening and light shift effect, as is underlined in Fig. \ref{fig:Rdependence}a, the index $R$ in the MFS mode can be 200 times larger than that in the FID mode (4000:20).
In the inset graph, the fluctuation of the index $R$ of FID mode, when the pump light power is less than 0.2 mW, may be due to the inadequate pumping. When the pump light power is low, the pumping rate is comparable to the atomic relaxation rate, and the atomic polarization of each hyperfine level is deficient, which refers to substandard SNR for fitting in data analysis.
When the pump light power is large, the index $R$ fluctuates within $\pm 4\times 10^{-7}$.

It is important to note that, even the AOM is not driven, the first order light exists. Whereas the extinction ratio of the AOM applied to modulate the pump light is not given, we assume it to be the typical value at -40 dB. When the peak value of the first order light power is 3.75 mW when the AOM is driven, the power of the first order light is 0.375 $\rm \mu W$ when the AOM is not driven. The corresponding light intensity $I_{\rm 1st,0}$ is given by
\begin{equation}
I_{\rm 1st,0}=(0.375~{\rm \mu W})/(\rm 4~mm^2)\approx0.01~{\rm mW/cm^2},
\end{equation}
which is much less than the saturation intensity for Cs D1 transition 1.67 $\rm mW/cm^2$ \cite{Cs}, which means there is almost no pumping effect because the pumping rate is much less than the atomic relaxation rate. 
Furthermore, given that the pump light frequency is tuned to the center of the Doppler-broadened Cs D1 $F_g = 3 \rightarrow F_e = 3$ resonance, the circularly polarized pump light should not bring vector light shift effect in theory, which is given by \cite{Hu2018}
\begin{equation}
\begin{aligned}
\Delta f_{\rm VLS}=&-\frac{\langle|E_0^2|\rangle}{2}\sqrt{\frac{6F(2F+1)}{(F+1)}}\epsilon
\sin\phi\\
&\times\frac{m_F}{2F}|\langle 6S_{1/2}|er|6P_{1/2}\rangle|^2\mathcal{F}(\nu),
\label{eq:vls}
\end{aligned}
\end{equation}
where  $\langle E_0^2\rangle$ is the average optical electric field experienced by atoms, proportional to the light intensity as $\langle E_0^2\rangle=2I_0/c\varepsilon_0$ where $I_0$ is the light intensity and $\varepsilon_0$ is the permittivity of vacuum, $\epsilon=1$ is the ellipticity of the left circularly polarized pump light, $\phi$ is the deviation of the pump light direction from orthogonality to the magnetic field (in our system $\phi\lesssim 1\times 10^{-4}~\rm rad$), $\langle 6S_{1/2}|er|6P_{1/2}\rangle\approx 3.17ea_0$ is the transition dipole matrix element between the $6S_{1/2}$ and $6P_{1/2}$ states, and $\mathcal{F}(\nu)$ is sum of the imaginary part of three Voigt profiles \cite{Happer1967,Costanzo2016} with dimension $\rm Hz^{-1}$ as
\begin{equation}
\mathcal{F}(\nu)=\sum_{F_e}\begin{Bmatrix}1 & 1 & 1\\F & F & F_e\end{Bmatrix}\begin{Bmatrix}J & J_e & 1\\F_e & F & I\end{Bmatrix}^{2}(2F_e+1)\Im[\mathcal{V}(\nu)],
\label{Fdefine}
\end{equation}
where $F_e$ represents the hyperfine levels in $6P_{3/2}$, $J_e=1/2$ for D1 transition, $I=7/2$ for Cs and $\mathcal{V}(\nu)$ is the Voigt profile corresponding to the interactions between atoms and light with different frequency $\nu$. 
Taking the fluctuation of pump light frequency ($\sim$20 MHz within one day) into consideration, the vector light shift for MR signals frequencies are
\begin{equation}
\begin{aligned}
\Delta f_{3,\rm{pump,VLS}}&\approx -1.03\times10^{-3}~{\rm Hz},\\
\Delta f_{4,\rm{pump,VLS}}&< 10^{-6}~{\rm Hz}, 
\end{aligned}
\label{pumpvls}
\end{equation}
The frequency shift for $F_g=4$ is negligible because the pump light is $\sim 9.2~\rm GHz$ detuned from any $F_g=4\rightarrow F_e$ transition.
The corresponding upper limit of systematic errors can be calculated by Eq. (\ref{eq:Rerror}) as
\begin{equation}
R_{\rm err,pump,VLS}\lesssim 8.5\times 10^{-8}.
\label{eq:pumperror}
\end{equation}
The systematic errors due to vector light shift from the pump light can be further suppressed by stabilizing the pump light frequency.

Improper pump light modulation frequency is another reason for prominent frequency shift in MR signals \cite{Swallows2013}.
This effect comes from the asynchronous optical pumping.
If the pump modulation frequency is detuned from the spin precession frequency, the spin polarization will be tipped along the applied magnetic field and precesses around the fictitious magnetic field of the vector light shift induced by the pump light \cite{Kimball2017}.
The fictitious magnetic field due to vector light shift is given by
\begin{equation}
B_{F,\rm LS}=\frac{\Delta f_{F,\rm VLS}}{\gamma_F}.
\end{equation}
The fictitious magnetic field generated by the pump light in Step 1 of Fig. \ref{fig:setup}b are calculated to be
\begin{equation}
B_{3, \rm LS}=2.94~{\rm nT},~~~~~B_{4, \rm LS}\sim 0.
\end{equation}
The fictitious magnetic field modulated at frequency $f_{\rm mod}$ can be treated as an magnetic field rotating at frequency $f_{\rm mod}$ transverse to the real magnetic field generated by the coils.
If the modulation frequency does not match the spin precession frequency, the phase of the spin precession signal will be shifted at $\delta\varphi$.
And the frequency shift $\Delta f_{\rm asyn}$ due to asynchronous pumping is the given by the phase shift of the spin precession signal over pumping time, taking the form as \cite{Swallows2013}
\begin{equation}
\Delta f_{\rm asyn}=\frac{\delta\varphi}{t}=\frac{1}{t}\frac{P_{y'}}{P_{z'}}=\frac{1}{t}\frac{\delta f\cdot f_{\rm LS}}{(\delta f)^2+(\nu_{F})^2},
\label{eq:asynerr}
\end{equation}
where $P_{y'},P_{z'}$ is the polarization along axis of the rotating frame, with the expressions already given in Ref. \cite{Budker2013}, $\delta f$ is the difference between the modulation frequency and the spin precession frequency, $f_{\rm LS}$ is the vector light shift from the pump light, $t$ is the total time of the pump light, and $\nu_F$ is the MR width (HWHM) of the hyperfine level. 
In our system, the time consumed to polarize atoms in one hyperfine level is 0.5 s, and the duty cycle of the modulation is 20$~\%$, so the total time of the pump light is $t=0.1~\rm s$.
Because the vector light shift in $F_g=4$ is negligible, only $F_g=3$ will be affected by the asynchronous pumping.
As is shown in Fig. \ref{fig:power}c, the MR width of $F_g=3$ when the pump light is 0.75 mW (on average after AOM) is $\nu_3=33~\rm Hz$.
To minimize the systematic errors from asynchronous optical pumping, in our system, the pump light modulation frequencies are tuned to within $\delta f\lesssim 2$ mHz of $\omega_{4}$ and $\omega_{3}$ during Step 1 \& 2 in Fig. \ref{fig:setup}b.
Substituting the parameters in our system into Eq. (\ref{eq:asynerr}), the frequency shift for $F_g=3$ is calculated to be:
\begin{equation}
\Delta f_{3,\rm asyn}=\pm 1.892\times 10^{-4}~\rm Hz.
\end{equation}
and the upper limit of the overall systematic errors of $R$ from asynchronous optical pumping is
\begin{equation}
R_{\rm err,asyn}= 1.564\times 10^{-8}.
\label{eq:asynerror}
\end{equation}

\subsubsection{Probe light}

The systematic errors from the probe light are insignificant in the Cs comagnetometer.
In general, the probe light may induce light shift and power broadening, same with the pump light.
Besides, the alignments generated by a linearly polarized light may be converted to orientations by external interactions, such as the magnetic field gradients, anisotropic collisions or electric fields \cite{Auzinsh2015}.

The light shift effect induced by the probe light, can be analyzed in two parts: the vector element and the tensor element.
The vector light shift is proportional to the product of the ellipticity $\epsilon$ and the deviation for propagation direction the probe light from orthogonality to the magnetic field  $\theta$ \cite{Kimball2013}. 
In our system, the ellipticity of the linearly polarized probe light is $\epsilon\lesssim 1.6\times 10^{-4}~\rm rad$ after the isolator (extinction ratio 38~dB), and the deviation is measured to be $\theta\lesssim 1\times 10^{-4}~\rm rad$. 
For Cs D2 transition, the corresponding dipole matrix element in Eq. (\ref{eq:vls}) should be replaced as $\langle 6S_{1/2}|er|6P_{3/2}\rangle\approx 4.48ea_0$, and the $J_e=1/2$ should be replaced by $J_e=3/2$. The vector light shifts due to the probe light for atomic spin precession frequencies are
\begin{equation}
\begin{aligned}
\Delta f_{3,\rm{prb,VLS}}&\approx -1.14\times10^{-4}~{\rm Hz},\\
\Delta f_{4,\rm{prb,VLS}}&< 10^{-6}~{\rm Hz}.
\end{aligned}
\end{equation}
The corresponding upper limit of the systematic errors due to the vector light shift of the probe light is calculated to be
\begin{equation}
R_{\rm err,prb,VLS}=9.39\times 10^{-9}.
\end{equation}

The tensor light shift of Zeeman sublevel $|F,m_F\rangle$ induced by the linearly polarized light near Cs D2 line is given by \cite{Peck2016}:
\begin{equation}
\begin{aligned}
\Delta f_{\rm TLS}=&-\frac{\langle|E_0^2|\rangle}{2}\sqrt{\frac{40F(2F+1)(2F-1)}{3(F+1)(2F+3)}}
(3\cos^{2}\alpha-1)\\
&\times\frac{3m_{F}^{2}-F(F+1)}{2F(2F-1)}|\langle 6S_{1/2}|er|6P_{3/2}\rangle|^2\mathcal{T}(\nu),
\label{eq:tensorls}
\end{aligned}
\end{equation}
where $\alpha$ is the angle between the magnetic field and the probe light polarization, and $\mathcal{T}(\nu)$ is sum of the imaginary part of three Voigt profiles \cite{Happer1967,Costanzo2016} with dimension $\rm Hz^{-1}$, given by
\begin{equation}
\mathcal{T}(\nu)=\sum_{F_e}\begin{Bmatrix}1 & 1 & 2\\F & F & F_e\end{Bmatrix}\begin{Bmatrix}J & J_e & 1\\F_e & F & I\end{Bmatrix}^{2}(2F_e+1)\Im[\mathcal{V}(\nu)],
\label{adefine}
\end{equation}
where $F_e$ represents the hyperfine levels in $6P_{3/2}$, $J_e=3/2$ for D2 transition.
As is shown in Eq. (\ref{eq:tensorls}), Zeeman sublevels with $m_F=\pm F$ have the max tensor light shift.
In our system, the probe light polarization is fixed at $\alpha\approx 54.74^{\circ}$ to make the item $(3\cos^{2}\alpha-1)\approx0$.
The errors of $\alpha$ are $\Delta\alpha\lesssim 10^{-5}~\rm rad$ with
\begin{equation}
\Delta(3\cos^2\alpha-1)=-3\sin2\alpha\Delta\alpha\approx 2.8\times 10^{-5}~\rm rad. 
\label{eq:alphaerror}
\end{equation}
Combine Eq. (\ref{eq:tensorls}), Eq. (\ref{adefine}) and Eq. (\ref{eq:alphaerror}) with the probe light intensity and frequency detuning, and the max systematic errors induced by tensor light shift in measuring the frequencies are 
\begin{equation}
\begin{aligned}
\Delta f_{3,\rm prb,TLS}&\approx0.0036 \rm ~Hz,\\
\Delta f_{4,\rm prb,TLS}&<10^{-6}~\rm Hz.
\end{aligned}
\end{equation}
The negligible frequency shift for MR signal of $F_g=4$ comes from the fact that the probe light is 14~GHz blue-detuned from the center frequency of the Doppler-broadened Cs D2 $F_g=4\rightarrow F_e$ resonance. The upper limit of systematic errors due to the tensor light shift of the probe light is calculated to be
\begin{equation}
R_{\rm err,prb,TLS}=2.95\times 10^{-7}.
\end{equation}
The dependence of the index $R$ on the probe light frequency detuning is shown in Fig. \ref{fig:Rdependence}b. When the probe light is near resonant to Cs D2 $F_g=3\rightarrow F_e$ transitions, the errors are as large as $10^{-4}$. As the inset shows, when the probe light frequency detuning is $5.0\pm0.5~\rm{GHz}$ from Cs D2 $F_g=3\rightarrow F_e=4$ transition, the index $R$ fluctuates within $\pm 4\times10^{-7}$.

The power broadening induced by the probe light comes from the process that atoms absorb the probe light, and the broadening can be calculated from the absorption rate \cite{Seltzer2008}
\begin{equation}
\Gamma_{\rm abs}=\sum\sigma(\nu)\Phi(\nu),~~\sigma(\nu)\propto\Re[\mathcal{V}(\nu)],
\label{eq:absorptionrate}
\end{equation}
where $\Phi(\nu)$ is the total flux of photons of frequency $\nu$, $\sigma(\nu)$ is the the photon absorption cross-section determined by the real part of the Voigt profile $\Re[\mathcal{V}(\nu)]$ corresponding to the absorption of light with different frequency for atoms.
In our system, the probe light is 5~GHz blue-detuned to the center frequency of the Doppler-broadened Cs D2 $F_g=3\rightarrow F_e=4$ resonance, and the probe light absorption rates for atoms in both hyperfine levels $F_g=3~\&~4$ are more than 2 orders in magnitude slower than relaxation rate.
Therefore, the power broadening in the widths of MR signals from the probe light is negligible.
Moreover, because of the inappreciable probe light absorption rate, there is no alignment formed in the atomic ensemble.
Consequently, the alignment-to-orientation conversion is out of consideration in our comagnetometer system.
The dependence of the index $R$ on the probe light power is shown in Fig. \ref{fig:Rdependence}c. When the probe light power is lower than $\sim 1~\rm mW$, the errors are slightly larger, maybe because of the fitting errors arising from the small amplitudes of the measured rotation signals. When the probe power is larger, the index $R$ fluctuates within $\pm 4\times 10^{-7}$.

Compared with the single-beam scheme \cite{Wang2018}, the pump-probe two-beam scheme is more beneficial to suppressing the systematic errors from laser fields, as Figure \ref{fig:Rdependence}b shows.
While in the pump-probe two-beam scheme, the probe light can be tuned off-resonant to suppress the systematic errors induced by laser fields when the pump light is on-resonance to polarize the atomic spins.

\subsubsection{Atomic collisions}
In our system, Cs atoms are contained in the cylinder vapor cell with anti-relaxation coatings in the inner surface. And collisions to the Cs atoms can shift the spin precession frequencies, and further bring systematic errors to the comagnetometer system.

The spin-exchange (SE) collisions between Cs atoms in different ground-state hyperfine levels $F_g=3~\&~4$ are one of the factors which may shift the spin precession frequencies. 
The frequency shift due to SE collisions is given by \cite{Happer1977}
\begin{equation}
\Delta f_{F,\rm SE}\approx-\frac{\Delta\nu_{F,\rm SE}}{18f_F}\left(1-\frac{1}{(2I+1)^2}\right)\left(1-\frac{4}{(2I+1)^2}\right),
\label{eq:seshift}
\end{equation}
where $\Delta\nu_{F,\rm SE}$ is the broadening induced by SE collisions in MR width which can be calculated by Eq. (\ref{eq:spinexch}), $f_F$ is the Larmor frequency of hyperfine level $F_g$, and the approximation is made when $\Delta\nu_{F,\rm SE}\ll f_F$.
In our experiments, the applied magnetic field is $B_0=3463.8~\rm nT$. Substitute the calculated results of $\Delta\nu_{F,\rm SE}$ in Eq. (\ref{eq:width}) and the Larmor frequency of $F_g=3~\&~4$ into Eq. (\ref{eq:seshift}), and the shifts due to SE collisions in spin precession frequencies of hyperfine levels are
\begin{equation}
\begin{aligned}
\Delta f_{3,\rm SE}&=-7.31\times 10^{-6}~\rm Hz,\\
\Delta f_{4,\rm SE}&=-1.10\times10^{-6}~\rm Hz.
\end{aligned}
\label{eq:sefreqshift}
\end{equation} 
The upper limit of systematic errors of the index $R$ induced by SE collisions is calculated by substituting Eq. (\ref{eq:sefreqshift}) into Eq. (\ref{eq:Rerror}) as
\begin{equation}
R_{\rm err,SE}=6.13\times10^{-10}.
\end{equation}

The collisions between Cs atoms and the inner surface of the vapor cell can produce quadrupolar splittings in MR signals \cite{Venema1992,Kimball2013}, especially for cylinder cell. For a cylinder cell with length $l$ and diameter $d$, the shift for coherence frequency of two adjacent Zeeman sublevels $|m_1\rangle\langle m_2|$ is given by \cite{Wu1990}
\begin{equation}
\Delta f_{\rm quad}=\frac{\bar{v}A\langle\vartheta\rangle}{2\pi\cdot 2h}(m_2^2-m_1^2)P_2(\cos\psi),
\label{eq:quad}
\end{equation}
where $\bar{v}$ is the average thermal velocity of Cs atoms ($\bar{v}=216.733\rm~m/s$ when at 22 $^\circ\rm C$), $\langle\vartheta\rangle$ is the mean twist angle per wall collision of a Cs atom, $P_2(x)=(3x^2-1)/2$ is the second-order Legendre polynomial, $\psi$ is the angle between the magnetic field and the axis of the cell, the cell asymmetry parameter is
\begin{equation}
A=\frac{d-l}{d+2l},
\end{equation}
and the characteristic cell dimension is
\begin{equation}
\frac{1}{h}=\frac{1}{2l}+\frac{1}{d}.
\end{equation}

As is depicted in Section \ref{sec2} and Fig. \ref{fig:setup}, the axis of the cylinder Cs atomic vapor cell is along $\hat{y}$, perpendicular to the magnetic field along $\hat{z}$ ($\psi=90^\circ$), with length $l=2.5~\rm cm$ and diameter $d=2.5~\rm cm$. The uncertainty of cell asymmetry parameter $\Delta A$ of the Cs cell we use in our experiments mainly comes from that we do not know exactly the inner dimension of the cell. We assume the uncertainty of $d-l$ to be $0.1\rm~mm$, and we have
\begin{equation} 
\Delta A=\frac{\rm 0.1~mm}{\rm 2.5~cm+2\times 2.5~cm}=0.00133.
\end{equation}
We have no knowledge about the mean twist angle per wall collision of a Cs atom $\langle\vartheta\rangle$, so we assume it to be no more than twice the result for Xe atom measured in Ref. \cite{Wu1990} as
\begin{equation}
|\langle\vartheta\rangle|\lesssim10^{-4}~\rm rad.
\end{equation}
By taking these parameters into Eq. (\ref{eq:quad}), the upper limit of the shift for coherence frequency of two adjacent Zeeman sublevels takes the form as
\begin{equation}
\Delta f_{\rm quad}=(m_2^2-m_1^2)\times 6.9\times 10^{-5}\rm~ Hz,
\end{equation}
Because of the atomic population in the steady state, which has been discussed in \ref{subsubsec1}, the shifts for MR signals of hyperfine levels $F_g=3~\&~4$ are approximately to be
\begin{equation}
\Delta f_{3,\rm quad}\approx\Delta f_{4,\rm quad}=3.45\times 10^{-4}\rm ~Hz,
\end{equation}
and the upper limit of systematic errors of the index $R$ induced by the quadrupole splitting is calculated to be
\begin{equation}
R_{\rm err,quad}=4.04\times 10^{-8}.
\end{equation}
\subsubsection{Earth rotation}
As is described in Ref. \cite{Kimball2013,Heckel2008}, the gyro-compass effect arising from the decoupling of atomic spins from Earth rotation will shift the spin precession frequencies of atoms in hyperfine levels $F_g=3~\&~4$.
And the shift is given by
\begin{equation}
\Delta f_{\rm gyro}=f_{\rm Earth}\cos\beta,
\end{equation}
where $f_{\rm Earth}=1/86400~\rm s=1.16\times10^{-5}~\rm Hz$, and $\beta$ is the angle between the applied magnetic field and the axis of Earth rotation.

In our experiments, $\beta$ is equal to $90^\circ$ plus the laboratory latitude when the applied magnetic field is parallel to the gravity.
The experiments are executed in Beijing ($\rm 116.20^\circ E, 40.25^\circ N$), and the frequency shift
in MR signals of hyperfine levels is
\begin{equation}
\Delta f_{3,\rm gyro}=\Delta f_{4,\rm gyro}= -7.48\times10^{-6}~\rm Hz.
\end{equation}
The upper limit of systematic errors of the index $R$ induced by the gyro-compass effect is calculated to be
\begin{equation}
R_{\rm err,gyro}=8.74\times 10^{-10}.
\end{equation}

\subsection{Detecting spin-gravity couplings}\label{subsec:2}
The single-species Cs atomic comagnetometer based on optical FID signal can find its place in fundamental physics research, especially the spin-gravity couplings \cite{Safronova2018,Kimball2013,Kimball2017,Wu2018,Wang2018}, because of the simplicity of Cs atomic structure, and the elimination of main systematic errors from the magnetic field gradients and the laser light field.

According to the calculation in reference \cite{Kimball2015}, if there exists the spin-gravity coupling, the gyro-gravitational ratios of hyperfine levels in ground state of Cs are given by
\begin{equation}
\begin{aligned}
\chi_{3}&=-\frac{1}{8}\chi_e+\frac{9}{8}\chi_N=-\frac{1}{8}\chi_e-\frac{1}{8}\chi_p,\\
\chi_{4}&=\frac{1}{8}\chi_e+\frac{7}{8}\chi_N=\frac{1}{8}\chi_e-\frac{7}{72}\chi_p,
\end{aligned}
\label{eq:chi}
\end{equation}
where $\chi_e,~\chi_N$ and $\chi_p$ refer to the gyro-gravitational ratios of electrons, nucleus and protons, respectively.
Taking the spin-gravity couplings into consideration, the total Hamiltonian for atoms in hyperfine level $F$ consists of the spin-magnetic couplings $H_B$ and spin-gravity couplings $H_g$, given by
\begin{equation}
H=H_B+H_g=\mu_B g_F B_0 m_F+\chi_F g \cos\theta m_F,
\label{eq:hamilton}
\end{equation}
where $\mu_B$ is the Bohr magneton, $g_F$ is the Land$\rm \acute{e}$ factor for hyperfine level $F$, $m_F$ is the magnetic quantum number of atoms, $g$ is the acceleration due to gravity, and $\theta$ is the angle between the magnetic field and the Earth's gravitational field.
The atomic spin polarization precession frequency of the two hyperfine levels will be
\begin{equation}
\begin{aligned}
f_{3}=|\mu_B g_3 m_3+\chi_3 g \cos\theta|=\gamma_{3}B_{0}-\chi_{3}g\cos\theta/h,\\
f_{4}=|\mu_B g_4 m_4+\chi_4 g \cos\theta|=\gamma_{4}B_{0}+\chi_{4}g\cos\theta/h,
\end{aligned}
\label{eq:freq}
\end{equation}
where $\gamma_{3,4}$ is the gyromagnetic ratio defined as $\gamma_{3,4}=|\mu_B g_F/h|$.
Because the spin precession directions of atoms in different ground-state hyperfine levels are opposite, we define the direction of atoms in $F=I+j$ as $+$ and in $F=I-j$ as $-$, and therefore the contribution to the overall spin precession frequency of spin-gravity interactions are opposite, which explains the different signs ($\pm$) of spin-gravity item in the right hand side of Eq. (\ref{eq:freq}).
When we reverse the magnetic field $B_0$, the angle between the magnetic field and the Earth's gravitational field will become $\theta\rightarrow\theta+\pi$.
Therefore, the index $R$ in the reversible magnetic field $B_0$ is given by
\begin{equation}
R(\pm)=\frac{\gamma_{3}B_{0}\mp\chi_{3}g\cos\theta/h}{\gamma_{4}B_{0}\pm\chi_{4}g\cos\theta/h}/\frac{\gamma_{3}}{\gamma_{4}}-1.
\label{eq:R}
\end{equation}

To measure the spin-gravity couplings for protons, we suppress the spin-magnetic couplings and the spin-gravity couplings for electrons in common mode by constructing the parameter
\begin{equation}
\Delta R\equiv R(+)-R(-)\approx\frac{4\chi_p g\cos\theta}{9\gamma_{3}B_0 h}.
\label{eq:deltaRchi}
\end{equation}
The approximation is made considering $\gamma B\gg\chi_{e}g,~\chi_{p}g$, and the nuclear magnetic moment is neglected due to $g_j\gg g_I$, thus the spin-gravity couplings for electrons are cancelled.
\begin{figure}
\centering
\resizebox{\columnwidth}{!}{%
\includegraphics{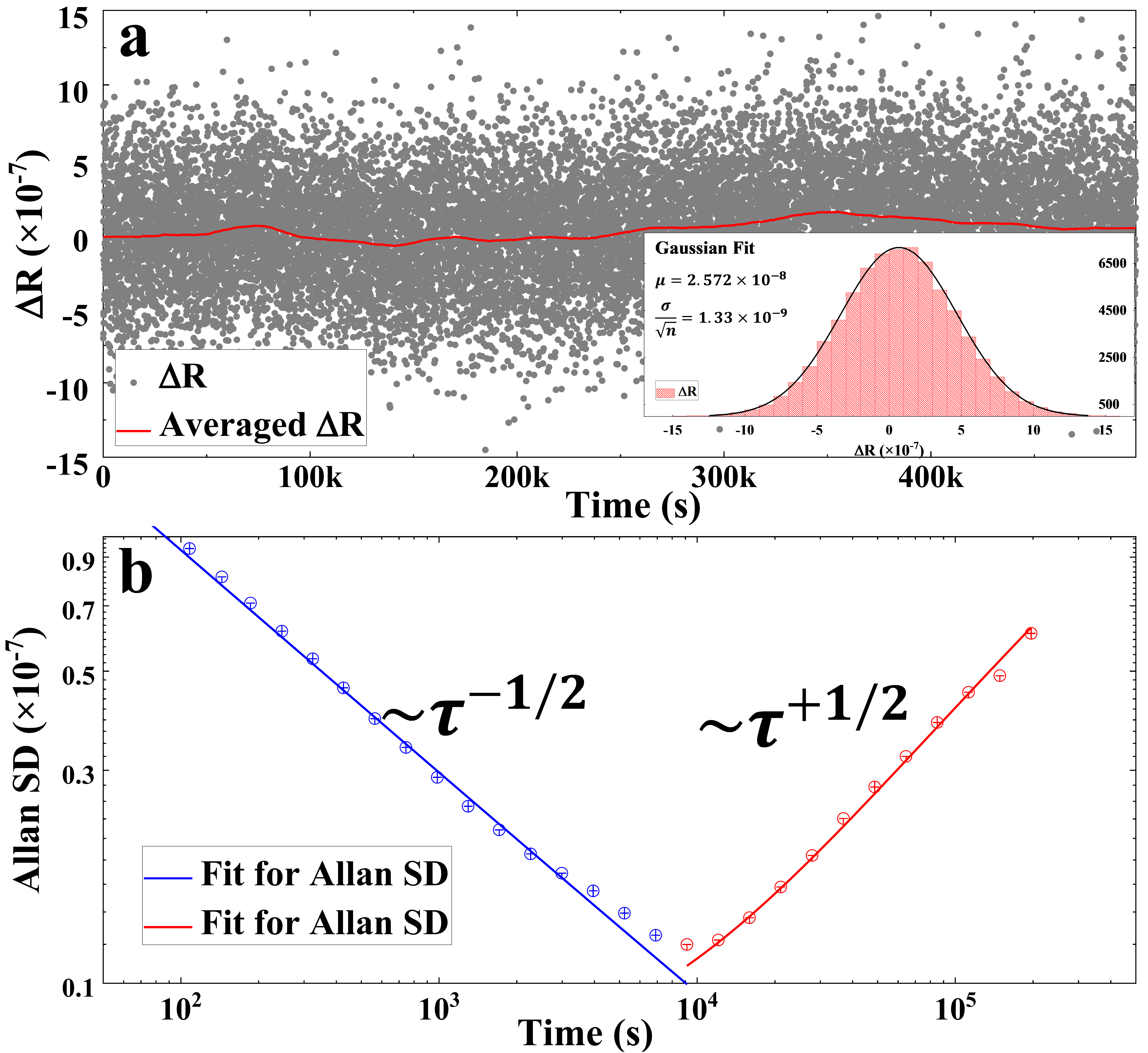}}
\caption{
\textbf{a} The statistic results of the parameter $\Delta R$ in a continuous five-day run. The results obey Gaussian distribution (mean value $2.572\times 10^{-8}$, standard error $1.33\times 10^{-9}$).
\textbf{b} The Allan standard deviation of the results in \textbf{a}. In the comagnetometer system, the applied magnetic field is (anti-)parallel to the gravity at $B_0=$3463.8 nT, and the $\Delta R$ can reach $10^{-8}$ at the averaging time of 10000 seconds.}
\label{fig:allan}
\end{figure}

In our comagnetometer system, the applied magnetic field is (anti-)parallel to the gravity at $B_0=$3463.8 nT.
One $\Delta R$ data is acquired by the successive measurement in $B(+)$ and $B(-)$. 
And the results of $\Delta R$ in a continuous five-day run is collected and shown in Fig. \ref{fig:allan}a.
The $\Delta R$ is between $\pm 1.5\times10^{-6}$ in time domain, and the average value of each 5000 runs fluctuates within $\pm 1.5\times10^{-7}$.
The long-term stability of the comagnetometer is indicated by Allan standard deviation (Allan SD, shown in Fig. \ref{fig:allan}b) and follows a $\tau^{-1/2}$ trend for $\tau < 10^4$ s, indicating a dominant white noise character and a stability at the $10^{-8}$ level. For $\tau>10^4$ s, the stability could be limited by the drifts in laser power and frequency, which introduce variations in the atomic population distribution within Zeeman sublevels after optical pumping.
The inset of Fig. \ref{fig:allan}a shows that the data obey Gaussian distribution with a mean value of $2.572\times10^{-8}$ and the standard error of mean value of $1.33\times10^{-9}$.

Effects that may induce systematic errors to the index $R$ are already discussed in \ref{analysis}, but things are slightly different for $\Delta R$.
The parameter $\Delta R$ is the difference of the index $R$ in opposite directions, and the systematic errors to the parameter $\Delta R$ is given by
\begin{equation}
(\Delta R)_{\rm err}=\sqrt{R_{\rm err}^2(+)+R_{\rm err}^2(-)}\approx\sqrt{2}R_{\rm err}.
\label{eq:DRerror}
\end{equation}
And as is discussed in Ref. \cite{Kimball2013}, in calculating the systematic for $\Delta R$, the systematic errors of the index $R$ from effects which are not related to the direction of the magnetic field will be suppressed by a factor of $\delta B/B$, the relative stability of the magnetic field. In our system, the factor is determined by the stability of the constant current source as $\rm 1~ppm~(10^{-6})$ within one day. 
Among all the effects discussed in \ref{analysis}, the effect due to asynchoronous optical pumping and the Earth rotation are related to the direction of the applied magnetic field.

Furthermore, the accuracy of inversing the applied magnetic field within the magnetic shield is another source of systematic errors for $\Delta R$.
Before the 5-day experiment begins, we have carefully calibrated the current to compensate the residual magnetic field within the shield and make $|B(+)|=|B(-)|$ as
\begin{equation}
|B(+)|=B_{\rm res}+k\cdot I_+=-B_{\rm res}+k\cdot I_-=|B(-)|,
\end{equation}
where $k$ is a constant of the coils with dimension $\rm nT/mA$, $B_{\rm res}$ is the initial residual magnetic field in the shield, and $I_{\pm}$ is the applied current.
But the change of geomagnetic field will alter the residual magnetic field within the shield, and the optical pumping will be ``desynchronized''.
The daily change of the geomagnetic field in Beijing is measured to be no more than 50 nT.
And the drift of the residual magnetic field $\Delta B_{\rm res}$ within the shield (shielding factor $10^{-5}$) is no more than
\begin{equation}
\Delta B_{\rm res}\leq 50\times10^{-5}=5\times10^{-4}~\rm nT.
\end{equation}
Then the difference of the atomic spin precession frequency in opposite magnetic fields is calculated to be $\Delta B_{\rm res}\times\gamma_F=3.5~\rm mHz$ for both hyperfine levels.
Substituting the frequency difference into Eq. (\ref{eq:asynerr}) and Eq. (\ref{eq:DRerror}), the upper limit of systematic errors for $\Delta R$ is
\begin{equation}
(\Delta R)_{\rm err,drift}=1.93\times 10^{-8}.
\end{equation}

\begin{table}
\caption{Estimated Upper Limits on the Effects of Various Sources of Systematic Errors on $\Delta R$}
\label{tab2}      
\centering
\begin{tabular}{lc}
\hline\noalign{\smallskip}
\textbf{Source of Systematic Errors}~~~~~~~~~ & \textbf{Effect on $\Delta R$}\\
\noalign{\smallskip}\hline\noalign{\smallskip}
Asynchronous optical pumping & $2.21\times 10^{-8}$\\
Drift of residual field & $1.93\times 10^{-8}$\\
Earth Rotation & $1.24\times 10^{-9}$\\
Tensor light shift & $4.17\times 10^{-13}$\\
Residual pump light & $1.20\times 10^{-13}$\\
Geometric phase effect & $7.52\times 10^{-14}$\\
Quadrupole splitting effect & $5.71\times 10^{-14}$\\
Nonlinear Zeeman effect & $3.01\times 10^{-14}$\\
Vector light shift & $1.33\times 10^{-14}$\\
Spin-exchange collisions & $8.67\times 10^{-16}$\\
\noalign{\smallskip}\hline
\end{tabular}
\end{table}

The systematic errors from potential effects on $\Delta R$ are shown in Table \ref{tab2}.
And the overall systematic errors are calculated to be
\begin{equation}
\Delta R_{\rm err,sys}=\sqrt{\sum{(\Delta R)^2_{\rm err}}}=2.937\times 10^{-8}.
\end{equation}
Based on these measurements and analysis, we find the parameter
\begin{equation}
\Delta R=(2.572_{\rm mean}\pm 0.133_{\rm stat}\pm 2.937_{\rm sys})\times 10^{-8},
\end{equation}
where the mean value and its standard error are from the results in Fig. \ref{fig:allan}, and the systematic errors are the sum of items shown in Table \ref{tab2}. The upper limits on $\Delta R$ are
\begin{equation}
\Delta R\leq 5.642\times 10^{-8}
\end{equation}
Substituting the range into Eq. (\ref{eq:deltaRchi}), we find the gyro-gravity constant for protons to be
\begin{equation}
\chi_p\leq 1.02\times 10^{-32}~{\rm g\cdot cm}.
\end{equation}
And the spin-gravitational energy of protons is given by $E_p=\chi_p\cdot g$, in our results
\begin{equation}
E_p\leq 6.3\times 10^{-18}~{\rm eV}.
\end{equation}

The upper limit for the spin-gravity couplings for protons is comparable to the most stringent existing constraint \cite{Kimball2017}, but further measures need to be taken to optimize the comagnetometer system.
As is shown in Table \ref{tab2}, the systematic errors from the light field and magnetic field gradients are highly suppressed in our system, with more than 3 orders in magnitude better than those in Ref. \cite{Kimball2017}.
However, the asynchronous optical pumping and the drift for residual magnetic field in the shield bring the most systematic errors to the comagnetometer.
A better magnetic shield can help to suppress the drift due to the daily variation of geomagnetic field.
And to stabilize the magnetic field in the shield actively is another way to effectively reduce the systematic errors induced by asynchronous optical pumping. 

\section{Conclusions and outlook}\label{sec:4}
In conclusion, an all-optical single-species FID Cs atomic comagnetometer is proposed and implemented, with the results underlining that the systematic errors from magnetic field gradients, and laser light field, can be suppressed.
In a 5-day continuous operation, the single-species Cs atomic comagnetometer is proved capable of probing the spin-gravitational energy of protons at a level of $10^{-18}$~eV, comparable to the most stringent existing constraint on the spin-gravity couplings \cite{Kimball2017}.

There are some optimizations for the atomic comagnetometer system to set a more stringent constraint on long-range spin-gravity couplings.
The systematic errors mainly come from the asynchronous optical pumping, which can cause transverse magnetism in the system and shift the spin precession frequency.
In the future, a better magnetic shield with active stabilization system will improve the stability of the magnetic field, and thus reduce the error between the modulation frequency for pump light and atomic spin precession frequency.

Additionally, in our system, the spin polarization is generated by synchronous optical pumping with circularly polarized light, which is tuned to the center of the Doppler-broadened Cs D1 $F_g=3\rightarrow F_e=3$ resonance.
The spin polarization in $F_g=4$ can be further strengthened by adopting another pump light tuned to the center of the Doppler-broadened $F_g=4\rightarrow F_e$ resonance, and meanwhile the spin polarization in $F_g=3$ will be also amplified due to the repopulation of the excited atoms from $F_g=4$.
The optimized spin polarization in both hyperfine levels will lead to a better SNR in the system, and improve the sensitivity of the comagnetometer by more than one order of magnitude.

Our comagnetometer is capable of meeting the demands for researches on searching for exotic spin-dependent interaction \cite{Safronova2018,Kimball2013,Kimball2017,Wu2018,Wu2019}.

\textbf{Acknowledgements} This work was supported by the National Natural Science Foundation of China (NSFC) (Grant Nos. 61571018, 61531003, 91436210), National Science Fund for Distinguished Young Scholars of China (61225003), and National Hi-Tech Research and Development (863) Program.
%
%

\end{document}